\documentclass[draft,jgrga]{agu2001}
\usepackage{graphicx}

\authorrunninghead{MELINI AND PIERSANTI}

\titlerunninghead{GLOBAL SEISMICITY AND SEALEVEL}

\authoraddr{D. Melini,
Istituto Nazionale di Geofisica e Vulcanologia, Via di Vigna Murata 605,
I-00143 Roma, Italy.
(melini@ingv.it)}
\begin{document}

%
%

%
\setkeys{Gin}{draft=false}

%
%

%
%

\title{Impact of global seismicity on sea level change assessment}

%
%


\author{D. Melini and A. Piersanti}
\affil{Istituto Nazionale di Geofisica e Vulcanologia, Rome, Italy}



%
%
%

%
%


\begin{abstract}
We analyze the effect of seismic activity on sealevel variations, by computing
the time-dependent vertical crustal movement and geoid change due to coseismic
deformations and postseismic relaxation effects. Seismic activity can
affect both the absolute sealevel, by changing the Earth gravity field and
hence the geoid height, and the relative sealevel, i.e. the radial distance
between seafloor and geoid level. By using comprehensive
seismic catalogues we assess the net effect of seismicity on tidal
relative sealevel measurements
as well as on the global oceanic surfaces,
and we obtain an estimate of absolute sealevel variations of seismic origin.

We improved the computational methods adopted in previous analyses 
considering the issue of water volume conservation through the
application of the sealevel equation and 
enabling 
us to evaluate the effect of an extremely large number of earthquakes on large grids 
covering the whole oceanic surfaces. These new potentialities allow us to perform more 
detailed investigations discovering a quantitative explanation for the overall 
tendency of earthquakes to produce a positive global relative sealevel variation.
Our results confirm the finding of a previous analysis that, on a global scale, 
most of the signal is associated with few giant thrust events, 
and that RSL estimates obtained using tide-gauge data 
can be sensibly affected by the seismic driven sealevel signal.

The recent measures 
of sealevel obtained by satellite altimetry show a wide regional variation of 
sealevel trends over the oceanic surfaces, 
with the largest deviations 
from the mean trend occurring in tectonically active regions.
While our estimates of average absolute sealevel variations turn out to be 
orders of magnitude smaller than the satellite measured variations, we can
still argue that mass redistribution associated with aseismic tectonic processes 
may contribute to the observed regional variability of sealevel variations. A
detailed study of these tectonic contributions is important to acquire
a complete understanding of the global sealevel variations and will be subject of
future investigations.


\end{abstract}

%
%

%

\begin{article}

\section{Introduction}

Current estimates of relative sealevel variations on a secular time-scale
based on tide-gauge measurements
indicate a uniform rise in the range $1.75\pm 0.55$ mm/yr 
\citep{douglas3,church05}.
The uncertainty on this figure depends on the particular subset of 
observations employed, on their scatter, and on the method used to
correct the data for vertical land movements due to glacial isostatic
adjustment (GIA) \citep{mitrdav95}. The main contributions to 
last decades sealevel rise come from thermal expansion of oceans due to
global warming and ocean mass change due to glaciers and ice sheet melting.

Recent observations from satellite altimetry \citep{cazenave04} over the decade
1993--2003 gave a larger rate (3.1 mm/yr after removing GIA effects)
and, by allowing for measurements of sealevel on the whole oceanic
surfaces, evidenced a strong nonuniform geographical distribution of sealevel
changes, with some regions exhibiting rates about 10 times greater than 
the global mean and some other regions where the trend was inverted 
and negative variations up to 15 mm/yr were detected.

Since seismic events alter the equilibrium state of the solid Earth
and perturbate its gravitational field, they are also likely to produce
sealevel variations. The perturbation of the Earth's gravity field due
to mass redistribution following a seismic event affects the geoid level
and it is therefore responsible for a variation in the absolute sealevel.
The vertical deformation of the seafloor, together with the geoid change,
produce also a relative sealevel change. Relative sealevel is directly
measured at tide-gauge stations, while absolute sealevel is measured by
satellite altimetric missions.

In a previous work \citep{rsl01}, 
we investigated the effect of global seismic activity on the 
observed relative sealevel variations, and found that great earthquakes
have an overall tendency to produce a sealevel rise, affecting the
measurements taken at those tide-gauge sites commonly employed for
sealevel rise monitoring. On a global scale, most of the signal is 
associated with few giant thrust events that, depending on the viscosity
of the asthenosphere, can induce a sealevel signal of at least 0.1 mm/yr.
This result has been obtained adopting the seismic catalogue considered by
\cite{marz02}, which contains $778$ shallow earthquakes 
(depth $\le 70$km) with magnitude $M \ge 7$, and includes events from the
\cite{ps92} compilation and the CMT catalogue \citep{cmt1}.

Estimates of sealevel rise coming from water volume increase due to ocean
warming give a rate of about 0.5 mm/yr while the rate due to mass increase
from ice melting is highly controversial and recent estimates range from less than 0.5
mm/yr to 1.5 mm/yr \citep{levitus00,milldoug}.    
Therefore, the average
contribution to RSL coming from seismic activity maybe comparable with
estimates of individual climatological factors and, in regions with strong
seismotectonical activity, may represent locally a major contribution to RSL.

In this work, we compute the seismic contribution to sealevel rise on
the whole oceanic surfaces with a self-consistent approach that
takes into account ocean volume conservation. We use as seismic
datasets both the catalogue considered
by \cite{marz02} and the CMT catalogue up to July 31, 2004.
The seismic dataset by \cite{marz02} covers a longer time window,
which is a crucial feature in assessing long-term effects, while
the CMT catalogue has a shorter temporal coverage but it is
characterized by a lower magnitude threshold
and provides more reliable estimates of focal parameters.
We find that the global mean of sealevel trends induced by 
earthquakes is positive, but its geographical distribution is
higly variable, with a pattern that shows
several analogies with the satellite measurements 
\citep{nerem02,cazenave04}.

We also quantified the effect of CMT seismic activity on tide-gauge
measurements as we did in \cite{rsl01} for the \cite{marz02} catalogue.
While the CMT results qualitatively confirm the ones obtained previously,
their absolute value is much smaller, most likely because of the absence
of giant subduction events like Chile 1960 and Alaska 1964 in the CMT catalogue.

On December 26, 2004, an exceptionally large event stroke the Indonesian
region. According to current estimates, it is probably the second largest
event ever registered. We present here some preliminary result about its
effect on sealevel.

In order to gain better insight into the reasons for the global tendence
of seismicity to produce positive sealevel variations, we performed
a detailed synthetic analysis, investigating separately the contributions
to vertical displacements and geoid variations. We also computed separately
the RSL field induced by 1960 Chile and 1964 Alaska earthquakes and
found that these two earthquakes alone account for a large fraction
of the seismically-driven sealevel signal. Beyond this, we found that the 
reciprocal geometrical features of the relative sealevel signal associated with these two 
events and the distributions of the tide gauge stations is mostly responsible 
for the positive mean in the computed relative sealevel trend.

\section{Method}

\subsection{Postseismic deformation modeling}

To compute the time-dependent deformation and gravity field variations associated
with a seismic dislocation we adopted the model proposed by \cite{piersanti95},
a spherical model which assumes an incompressible,
layered, self-gravitating Earth with Maxwell linear viscoelastic rheology. The model
was later refined by \cite{soldati98} to account for gravitational
effects and by \cite{boschi00} to include the contribution of
deep (upper mantle) earthquakes. We refer the reader to these
works for details concerning the numerical approach.

We employed a 4-layer stratification which includes an $80$ km elastic lithosphere,
a $200$ km thick low-viscosity asthenosphere with $\eta = 10^{19}$ Pa s, 
appropriate for oceanic asthenosphere \citep{cf03}, a
uniform mantle with $\eta = 10^{21}$ Pa s and a fluid inviscid core.
All the other mechanical parameters have been obtained by means of a
weighted volume average of the corresponding parameters of PREM model.
 
Of course, our results depend on the chosen viscosity values.
The effect of varying asthenosphere and mantle viscosity on the
evolution of geophysical observables in the postseismic relaxation process has
been extensively discussed in a series of works 
\citep{piersanti95,piersanti97,soldati98,boschi00,nostro01} 
to which we refer the reader for more 
details. Roughly, we can say that viscosity values smaller than those employed
here would further enhance the postseismic contribution to deformation
and gravity fields, while the purely elastic coseismic
response would remain unchanged.

With this model, we computed the time-dependent deformation and gravity
field variations due to each seismic event in our catalogues.
The time-dependent relative sealevel $S(t)$ at a given observation site
is then computed from the vertical displacement $u_z$ and geoid variation $G$ as
$S(t) = G(t) - u_z(t)$.

The procedure to retrieve the numerical solution from our model involves 
a spherical harmonic expansion of the physical observables. 
Since the convergence of the harmonic series is very slow, it 
is needed to sum up to very high harmonic degrees ($1000\le l \le 8000$,
depending on source parameters), which requires huge computational resources.
Since we had to evaluate the contribution of over 20,000 earthquakes, we implemented
our codes on a parallel distributed-memory computer based on 48 Intel Xeon CPUs,
where the whole simulation procedure required about 2 months of CPU time to run.

\subsection{Conservation of water volume}
\label{water_volume}

The modeling approach described above allows us to estimate the sea 
level variation from the local changes in geoid height and seafloor
vertical displacements. However, when dealing with global modeling, we
should consider the problem of conservation of total water mass.
The most general approach to the problem of ocean water conservation
is expressed by the ``sea level equation'' (SLE), which takes into
account the contribution of geoid changes and vertical seafloor
displacements over the whole ocean \citep{douglas3, peltier98}. 
Moreover, SLE includes the immission of freshwater from melting of 
continental ice sheets (which are a major factor in postglacial sealevel 
variations) as well as the variation of the ocean function. 

In the present approach, the conservation of total water mass is a 
direct consequence of our formulation, since we do not admit
any water mass exchange; also, we don't include the effect of
global warming, so we can assume constant water density; from
these assumptions follows the conservation of total water volume.
Consequently, instead of using the most general form of SLE, we can use a simplified
approach. Let $u_r(\theta,\phi,t)$ and $G(\theta,\phi,t)$ be the 
vertical seafloor displacement and geoid change at coordinates $(\theta,\phi)$ 
and time $t$. We can express the total water volume conservation as:
\begin{equation}
\label{eq1}
\int_{\Omega'} \left( G(\theta,\phi,t)-u_r(\theta,\phi,t) \right) 
r_T^2 d\Omega = 0
\end{equation}
where $r_T$ is the Earth radius and the integration is carried out over
the surface of the oceans. Since we are dealing with very small sealevel
variations, we can safely neglect the variations of the ocean function
and assume the integration domain $\Omega'$ to be constant.

In our modeling approach the conservation of water volume is not
automatically guaranteed, so we have to correct our computations to
assure the validity of equation \ref{eq1}. To this aim, if $G_0$ is
the geoid change associated with seismic activity, we can introduce
a correction $G_1$ so that the total geoid variation $G=G_0+G_1$
satisfies equation \ref{eq1}. Since we are dealing with small
sealevel variations, at first order we can assume that this offset is 
constant over the oceanic surface, so that from equation \ref{eq1}
follows
\begin{equation}
\label{eq2}
G_1(t) = - \frac{
\int_{\Omega'} \left( G_0(\theta,\phi,t)-u_r(\theta,\phi,t) \right) r_T^2 
d\Omega}{ \int_{\Omega'} r_T^2 d\Omega }
\end{equation}
and we can write the relative sealevel at a given observation site as
$ S(t) = G_0(t) - u_r(t) + G_1(t) $.

The computation of the correction term in eq. \ref{eq2} involves the 
numerical integration of $G_0(\theta,\phi,t)$ and $u_r(\theta,\phi,t)$
over the oceanic surface.
Since the deformation field of earthquakes has very strong spatial 
variations in the near field, to carry out the numerical integration 
a dense sampling of the integrand function is required. The point dislocation
approximation that we used in our simulations represents a further potential
error source in the evaluation of the integral, because it produces large,
unrealistic values of the fields near the source. 
Beyond this, Since we are dealing
with simulations involving over 20,000 seismic events, we found that,
even adopting a massive parallel computing approach, in order to keep the simulation time
acceptable (of the order of one month of CPU time) we cannot perform the 
integration of eq. \ref{eq2} for the whole
seismic dataset. We considered instead a set of 8 extremely large earthquakes
occurred in the Pacific area, which account for about 80\% of the total seiemic
moment release in the last century \citep{casar_remtrig1}. Since most of
the seismic sealevel variation signal is associated with the largest events,
we computed the cumulative $G_1(t)$ with these 8 events and used it to
correct the sealevel time histories computed with the rest of the seismic
dataset.

\begin{figure}
\begin{center}
\noindent\includegraphics[width=40pc]{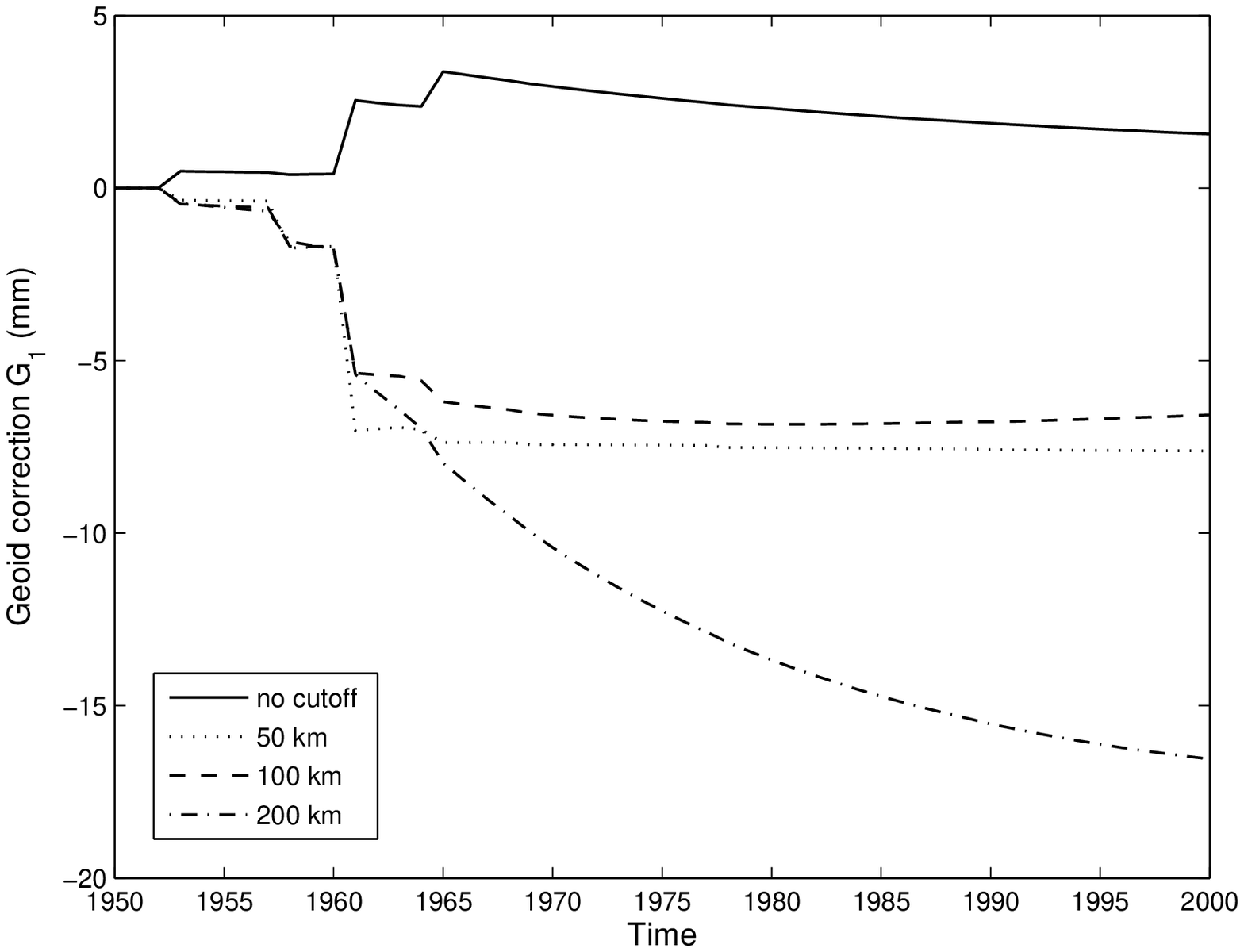}
\end{center}
\caption{Time-dependent geoid correction that accounts for global water volume
conservation. This correction has been computed using 8 earthquakes selected
between the largest ones occurred during the last century in the Pacific area.
The 4 different curves are computed excluding the integration points located
within 4 different cutoff values from the seismic sources
(cutoff values are 0, 50, 100 and 200km).}
\label{correz}
\end{figure}

To assess the effect of point source approximation in the near-field geoid
changes, we integrated eq. \ref{eq2} both using the whole integration domain
and excluding grid points located within a cutoff distance $d$ from the seismic
source. In figure \ref{correz} we compare the time-dependent $G_1(t)$ computed without 
cutoff and with $d=50,100,200$ km. From this figure we can infer that the 
correction due to water volume conservation 
is strongly dependent on the chosen cutoff value, confirming that
near-field effects give large contributions to $G_1$. Since the application
of an integration cutoff strongly affects the final correction term and can shadow
the physical signal, we decided to use the correction without cutoff to avoid the
introduction of any arbitrary (operator dependent) bias. We emphasize that this correction is
strongly dependent by the near-field signal, where we get ureliable values
of dislocation and gravity fields because of the point source representation,
and therefore $G_1$ will be undoubtly affected by approximation errors which 
may be quite large.

\subsection{Seismic datasets}

To compute the sealevel changes due to cumulative global seismicity 
we adopted two different catalogues. The first is the one considered by 
\cite{marz02}, and includes 778 shallow (depth $\le 70$ km) magnitude 
M $\ge 7$ earthquakes worldwide distributed in the period $1900-Ð1999$. This dataset
was compiled including data from the Centroid Moment Tensor (CMT) \citep{cmt1}
and \cite{ps92} catalogues; the focal parameters of the events
taken from the Pacheco and Sykes catalogue have been estimated by using the moment
tensor of the neighbor earthquakes reported by the CMT catalogue occurred within 
a certain distance from the Pacheco and Sykes epicenter, as explained by
\cite{marz02}. From now on, we will refer to this catalogue as ``PS''.

The second catalogue is the whole CMT catalogue, from 
January 1, 1976 up to July 31, 2004 that contains 21,708 events 
with magnitude M $\ge 4.7$. 

Basically, the PS catalogue covers a longer time window and includes 
the greatest events of the last century while the CMT catalog contains 
much more earthquakes with more reliable focal parameters, but it is 
characterized in average by much less energetic events.
As a consequence, we could roughly say  that the CMT catalog gives more 
precise information on the small scale features of the temporal evolution of
RSL signal at each site while the PS catalog give more realistic results
about the absolute magnitude of the seismic driven RSL signal.

\section{Results}

In our previous investigations \citep{rsl01}, we computed sealevel variations
of seismic origin only at the sites of PSMSL tide-gauge stations, also because
the solution of the numerical model on a large number of points, such as a grid
spanning the oceanic surfaces, had too heavy computational requests. For this work,
we developed a further optimized version of the numerical code exploiting a 
massive parallelism, that enables us to compute the relative sealevel variations 
on the whole oceanic surfaces and to use, at the same time, a much larger number
of seismic sources, such as the ones reported by the CMT catalogue. 

\subsection{Global effects}

\begin{figure}
\begin{center}
\noindent\includegraphics[width=30pc]{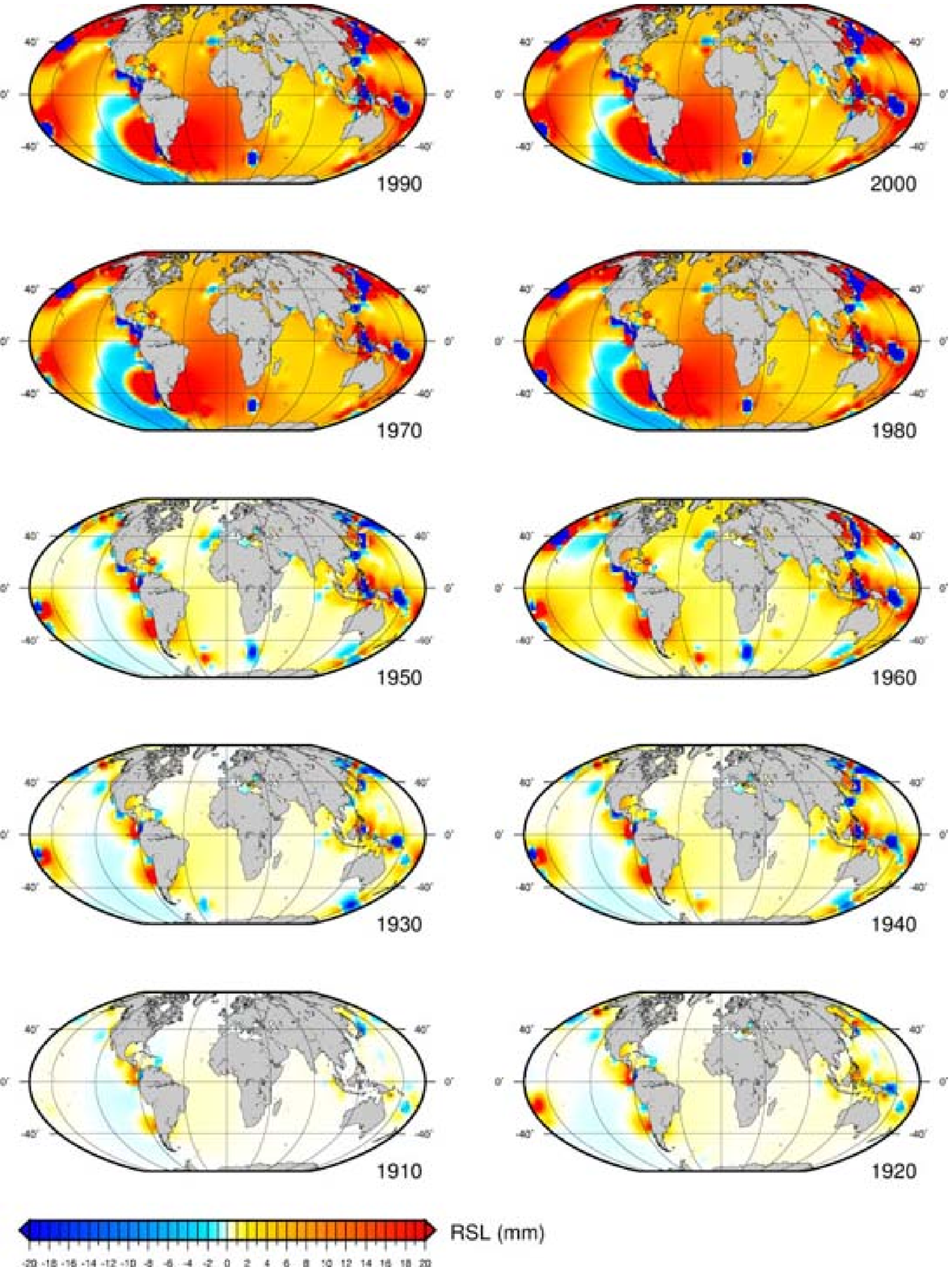}
\end{center}
\caption{Magnitude of relative sealevel variation over the oceans
computed at 10 different times resulting from PS seismic
activity. Each map includes the effect 
of all the earthquakes occurred since the beginning of the
catalogue.
}
\label{oceani_ps}
\end{figure}

\begin{figure}
\begin{center}
\noindent\includegraphics[width=30pc]{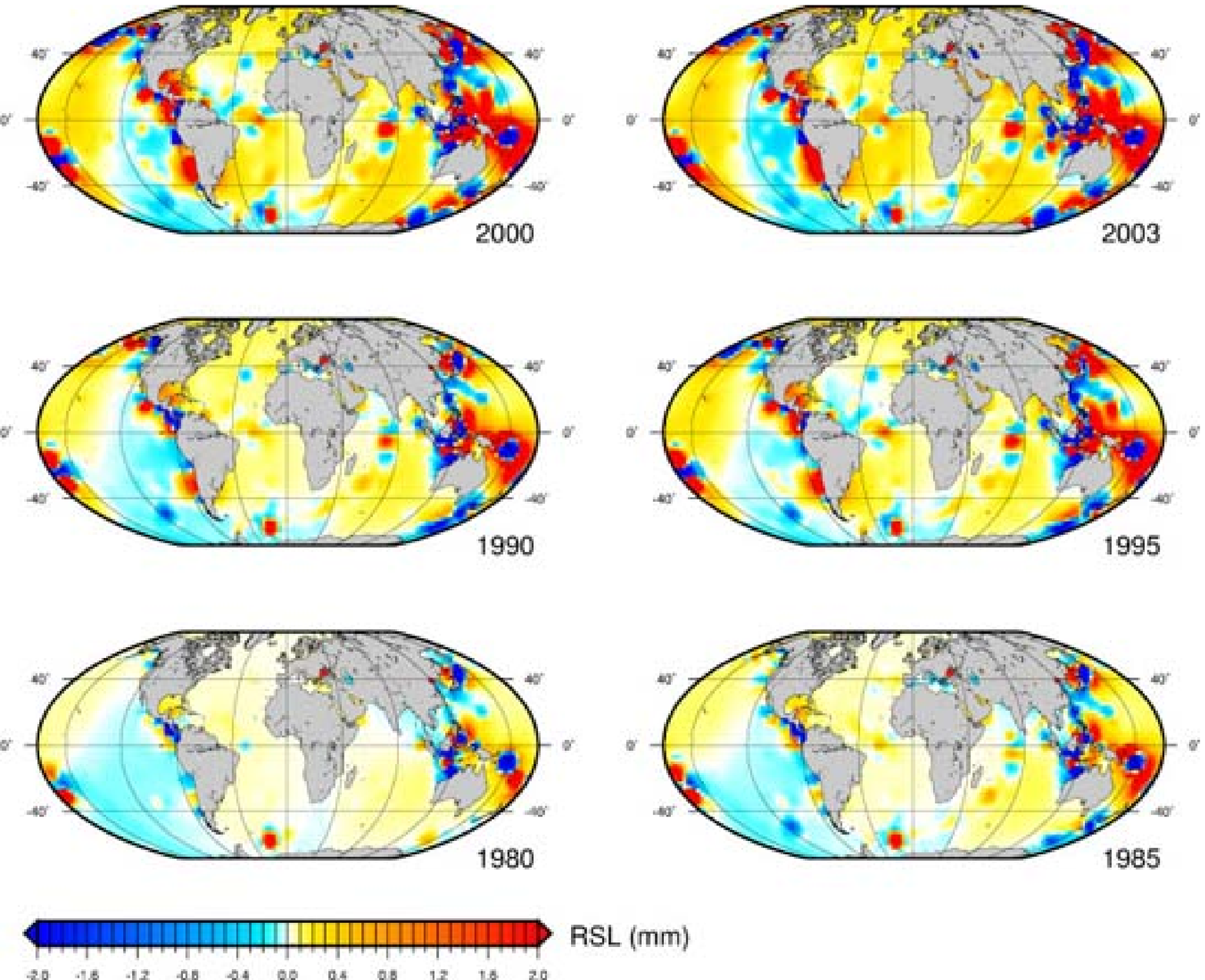}
\end{center}
\caption{Magnitude of relative sealevel variation over the oceans
computed at 6 different times resulting from CMT seismic
activity. Each map includes the effect 
of all the earthquakes occurred since the beginning of the
catalogue.}

\label{oceani_cmt}
\end{figure}

Figures \ref{oceani_ps} and \ref{oceani_cmt} show the magnitude of 
relative sea level variations, in the whole oceanic surface, 
computed at different time steps, due to the net effect of all the
earthquakes contained in the PS and CMT catalogues, respectively.

The results obtained with PS catalogue, shown in figure \ref{oceani_ps},
have been corrected to account for the conservation of total water volume as 
described in section \ref{water_volume}. Since the greatest earthquakes of the last century 
used to compute the geoid correction are included only in the PS catalogue,
we didn't apply it to the results obtained with the CMT
catalogue.

Even from a quick look at figures \ref{oceani_ps} and \ref{oceani_cmt} we can
see that the net effect of seismicity is a global increase of sealevel and that the magnitude 
of the induced variations shows a highly spatial variability reaching in some regions
values of several centimeters. 
We see also that nearly all the coastlines are located within zones
of positive RSL variation, while the larger areas of negative RSL variation
are mostly placed far from the coasts. Since the vast majority of PSMSL tide-gauge
stations are located along the coastlines, we expect that when
we compute the effect of seismically-driven RSL on the PSMSL station sites
we will find a positive signal, as, in fact, we did in our previous analyses \citep{rsl01}.

By comparing figures \ref{oceani_ps} and \ref{oceani_cmt} we can see also
that the RSL effect of earthquakes contained in the CMT catalog, which lacks
the giant thrust events of the last century, turns out to be at least
one order of magnitude smaller than the effect of PS seismicity. 
In fact, when we look at the temporal variation of sealevels in figure
\ref{oceani_ps}, the largest contribution to sealevel variations almost
on the whole oceanic surface comes in the interval (1960--1970), where
we register the occurrence of the 1960 Chile and 1964 Alaska earthquakes. 
This is a confirmation of what we obtained when looking at the 
seismic RSL effects on the time-histories of tide-gauge 
measurements \citep{rsl01}, i.e. that earthquake--induced
sealevel variations are mainly due to the effect of a few big earthquakes
rather than the superposition of many small contributions.

The recent launch of altimetric satellite missions allowed to obtain independent
estimates of sealevel variations. The TOPEX/Poseidon mission \citep{nerem95,nerem97} measured
absolute geocentric sealevels along a ground track uniformly covering the oceanic
surfaces. The acquired dataset covers the last decade (1993--2003) and indicates
a mean sealevel rise of $3.1$ mm/yr after correcting for GIA \citep{cazenave04}.

In the same time window, our results give a global mean of order $10^{-3}$ mm/yr for
relative sealevel while the rate for absolute (i.e. geoid) changes are an order of 
magnitude smaller. The sealevel trends obtained by global measurements such the
ones carried out by TOPEX/Poseidon are, in fact, mainly due to thermal expansion of
ocean water in response to global warming. Though it is clear that the global seismic 
selalevel signal is by far too small to be compared with the total detected signal
nevertheless it represents only a facet of
the broader problem of global tectonic effects on sealevel. In this respect,
it is reasonable to consider that the RSL signal associated with seismic
events is higly correlated, and probably much smaller, with that associated
with the whole tectonic processes since they share a common physical origin.
Therefore, even if a direct comparison between seismic driven sealevel variations 
and measured trends is not possible, we remark that
the role played by the whole tectonic processes could be important. Also the fact that 
in the detected data the main deviations
from the mean sealevel trend are located in tectonically active regions, suggests that this topic
deserves further investigations.

\subsection{Effects on PSMSL sites}

\begin{figure}
\begin{center}
\noindent\includegraphics[width=25pc,angle=-90]{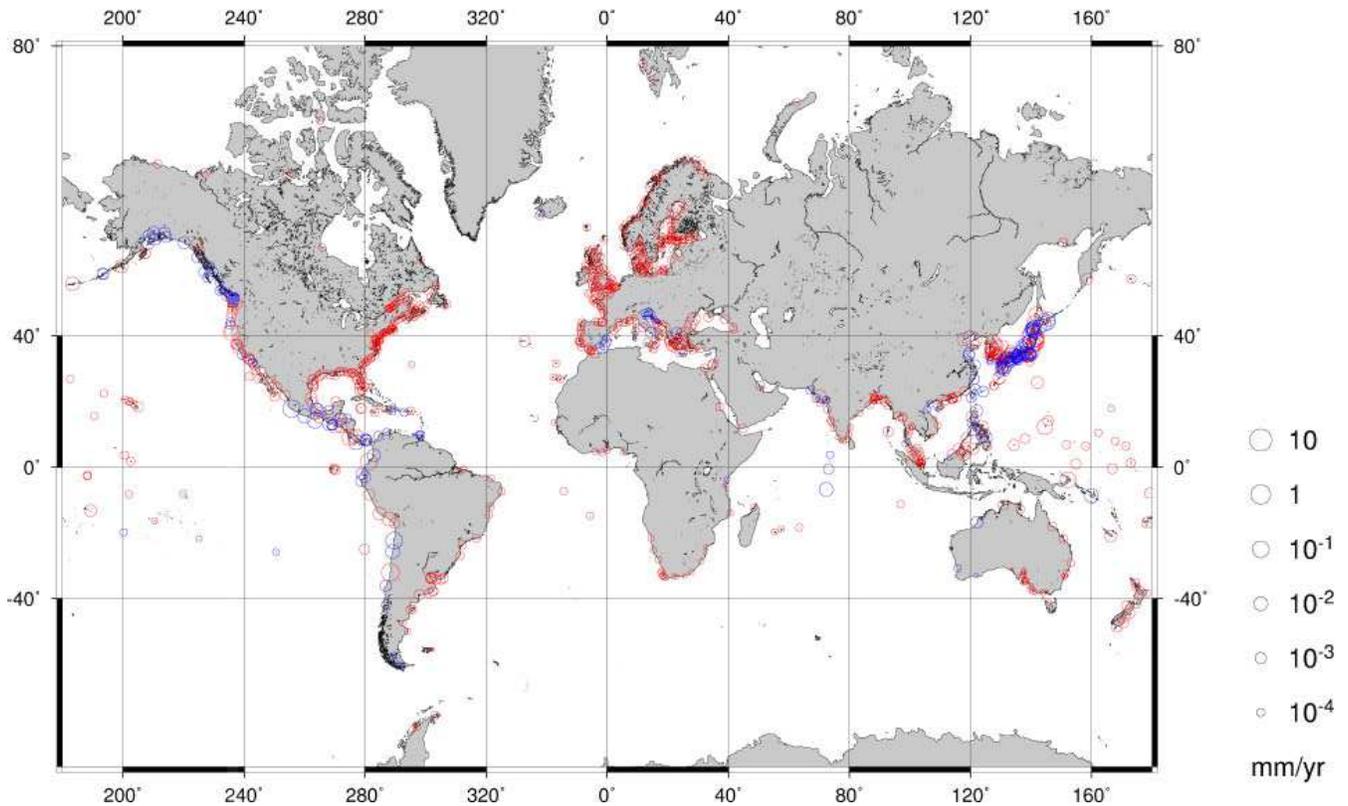}
\end{center}
\caption{Relative sealevel variation rates over 1976--2003 at the
locations of PSMSL tide-gauge stations, resulting from the cumulative
effect of CMT seismicity. Red and blue circles correspond to positive 
and negative trends, respectively.}
\label{pallocchi_cmt}
\end{figure}

\begin{figure}
\begin{center}
\noindent\includegraphics[width=25pc]{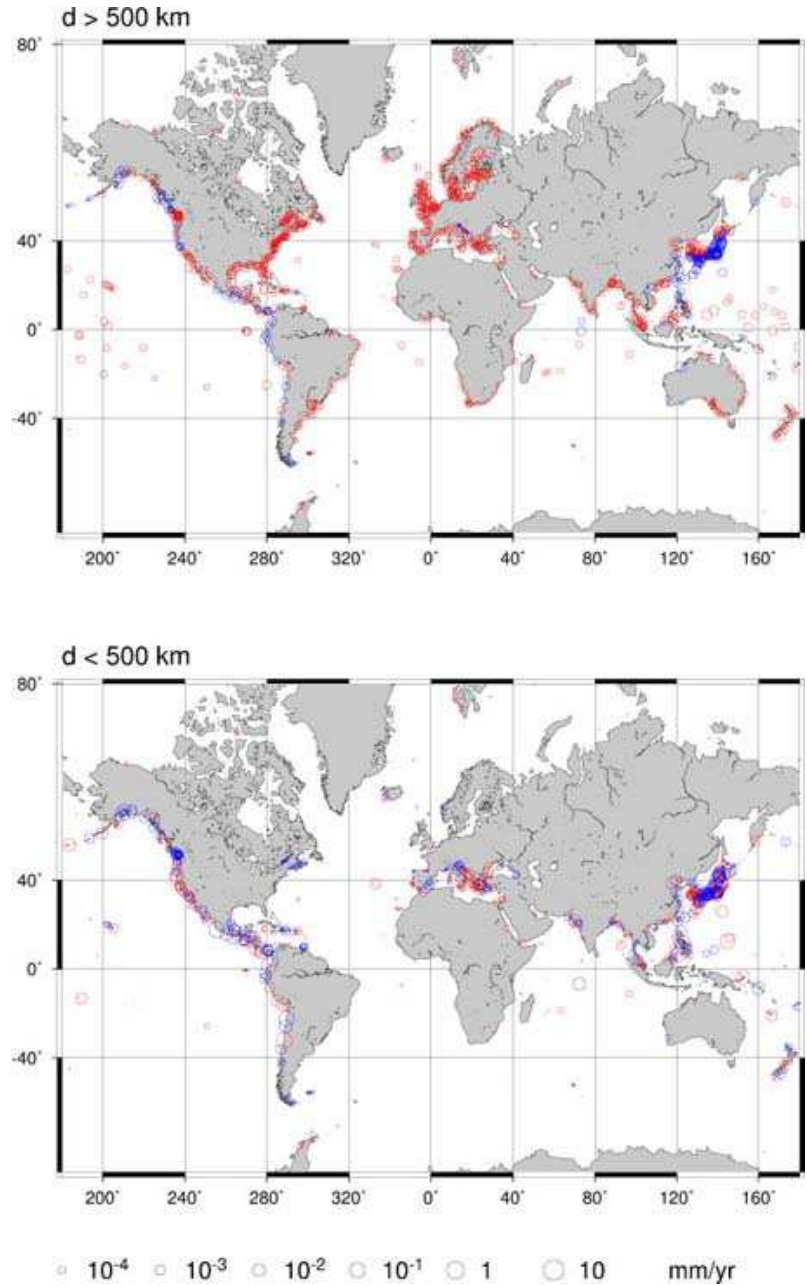}
\end{center}
\caption{Relative sealevel variation rates over 1976--2003 at the locations
of PSMSL tide-gauge stations, resulting from the cumulative effect
of CMT seismicity. In the lower panel it is shown only the contribution
of earthquakes located within 500 km from each tide-gauge, while in the upper
panel it is shown the contribution of earthquakes located at over 500 km
from the tide-gauge. Red and blue circles correspond to positive 
and negative trends, respectively.}
\label{pallocchi_cmt_nf}
\end{figure}

In figure \ref{pallocchi_cmt} we show the rate of earthquake-induced 
relative sealevel variations expected at each of the $1016$ PSMSL 
tide--gauge stations due to the cumulative effect of the CMT catalogue 
seismicity. The plotted values have been computed by least-squares
interpolation of the time-series $S(t)$ over (1976-2003); red and blue 
circles indicate sealevel rises and falls, respectively.
In figure \ref{pallocchi_cmt_nf} we plot the separate contribution to RSL variations due to 
near-field sources (distance from the tide-gauge within $500$ km) and far-field sources 
(distance is greater than $500$ km).

Also from the results of figure \ref{pallocchi_cmt} we obtain that the 
effect of seismicity on the RSL variations measured by tide-gauge
stations is a positive trend. According to our simulations, the
average rate $\sum_k S_k/N$ over all the 1016 PSMSL sites is 
0.019 mm/yr for the CMT catalogue (while for the PS catalogue we found 0.25 mm/yr).
Major deviations from the general trend are observed in the Mediterranean 
area and along the circumpacific ring (North American west coast, Chile and Peru). 

When we look at the RSL fields of figures \ref{oceani_ps} and \ref{oceani_cmt}
we can see that the areas where we obtain a negative variation are limited
in extension, except for the large negative lobe off the Pacific coast of South America,
that is associated with the 1960 Chile event. This behavior suggests that 
the negative sealevel variations are generally associated with the local effect
of relatively small events. Indeed, when we separate the contributions to
RSL measured by tide-gauges coming from ``near'' and ``far'' earthquakes (figure
\ref{pallocchi_cmt_nf}), we see that most of the negative contributions comes
from earthquakes located within 500 km from the tide-gauge.



\begin{figure}
\begin{center}
\noindent\includegraphics[width=30pc]{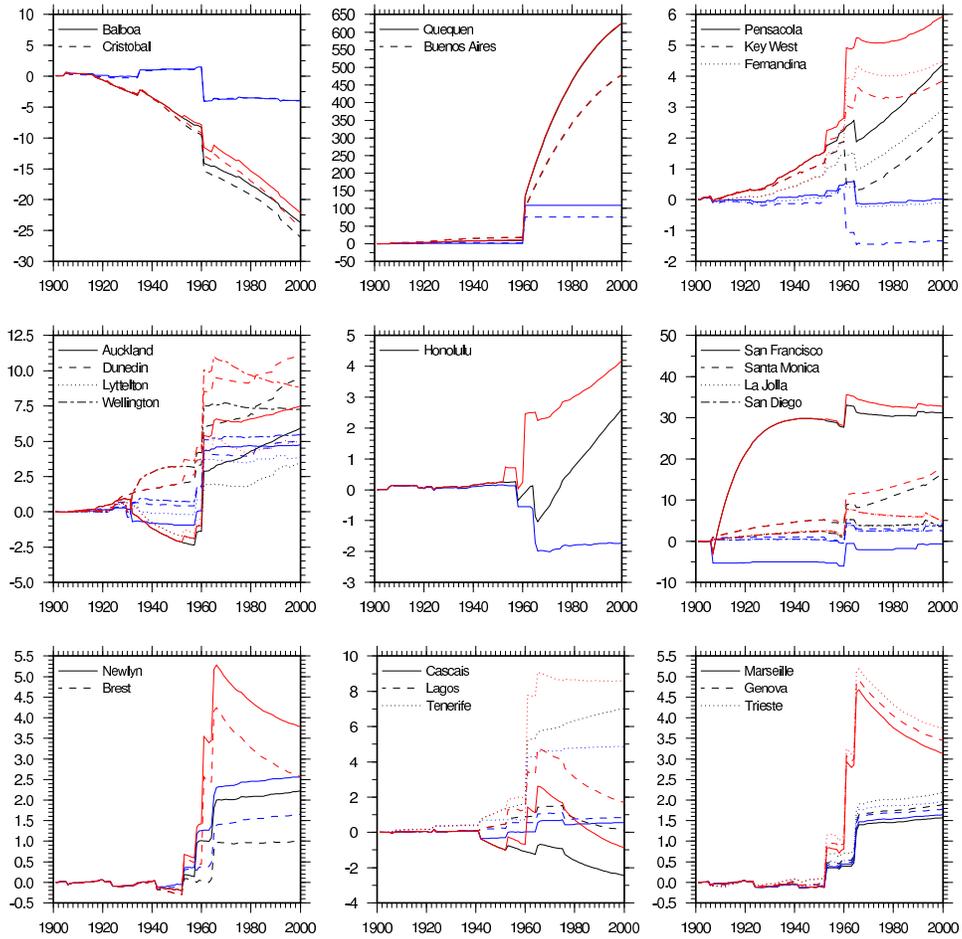}
\end{center}
\caption{Relative sealevel variation time-histories resulting from
PS seismicity on the the tide-gauge sites considered by \cite{douglas2}.
Black and blue lines represent the viscoelastic and purely elastic responses, without
taking into account the water volume conservation. Red lines represent the viscoelastic
response with water volume conservation.}
\label{douglas_ps}
\end{figure}

\begin{figure}
\begin{center}
\noindent\includegraphics[width=30pc]{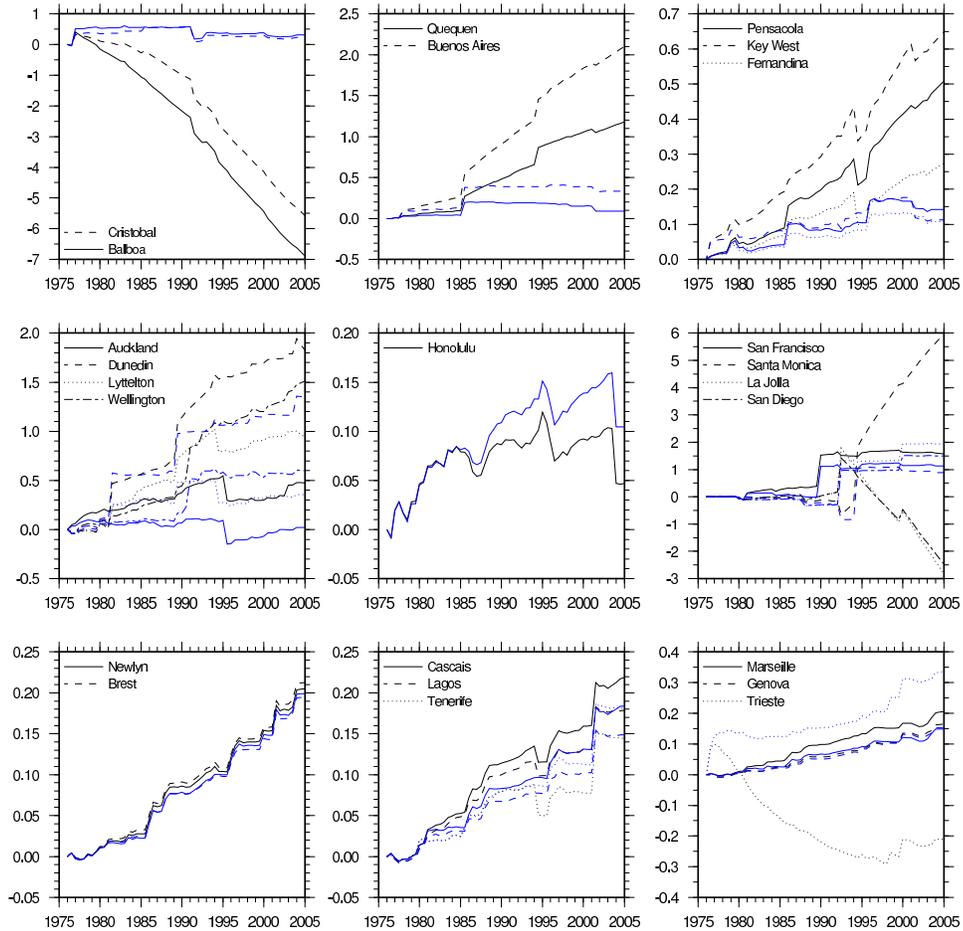}
\end{center}
\caption{Relative sealevel variation time-histories resulting from
CMT seismicity on the tide-gauge sites considered by \cite{douglas2}.
Black and blue lines represent the viscoelastic and purely elastic responses, respectively.}
\label{douglas_cmt}
\end{figure}

In figures \ref{douglas_ps} and \ref{douglas_cmt} we turn our attention
to the details of sealevel time-histories of the same PSMSL tide-gauge stations
which have been considered by \cite{douglas2} for his estimate of long-term sealevel rise. 
This set of 24 sites have been selected by Douglas for the length of sealevel records, 
which exceeds 70 years, for their expected tectonical stability and for their worldwide coverage. 
We have grouped these PSMSL sites regionally as done by \cite{douglas1,douglas2}. 

When we look at the effect of PS seismicity (figure \ref{douglas_ps}) we can see
that most of the seismic RSL signal is due to the coseismic effects of a few
giant earthquakes, mainly the 1960 Chile and 1964 Alaska earthquake. The postseismic
relaxation plays an important but not primary role, except for the stations
located near the epicenter of large events (i.e. San Francisco, Quequen, Buenos
Aires). However, in these cases the point source approximation used in our analysis
gives unreliable results and a more realistic computation, based on a finite size
seismic source, is needed.  On the other hand, in the RSL time--histories obtained from CMT seismicity (figure
\ref{douglas_cmt}) we see that the dominant effect comes from postseismic relaxation,
because of the absence of giant earthquakes in the time window covered by the catalogue.
This figure allows us also to evaluate the impact of the correction 
for water volume conservation on the total RSL signal; though not negligible, we can see that 
its effect becomes important only in those sites where modest values of seismic RSL signal are 
registered.

\begin{figure}
\begin{center}
\noindent\includegraphics[width=30pc]{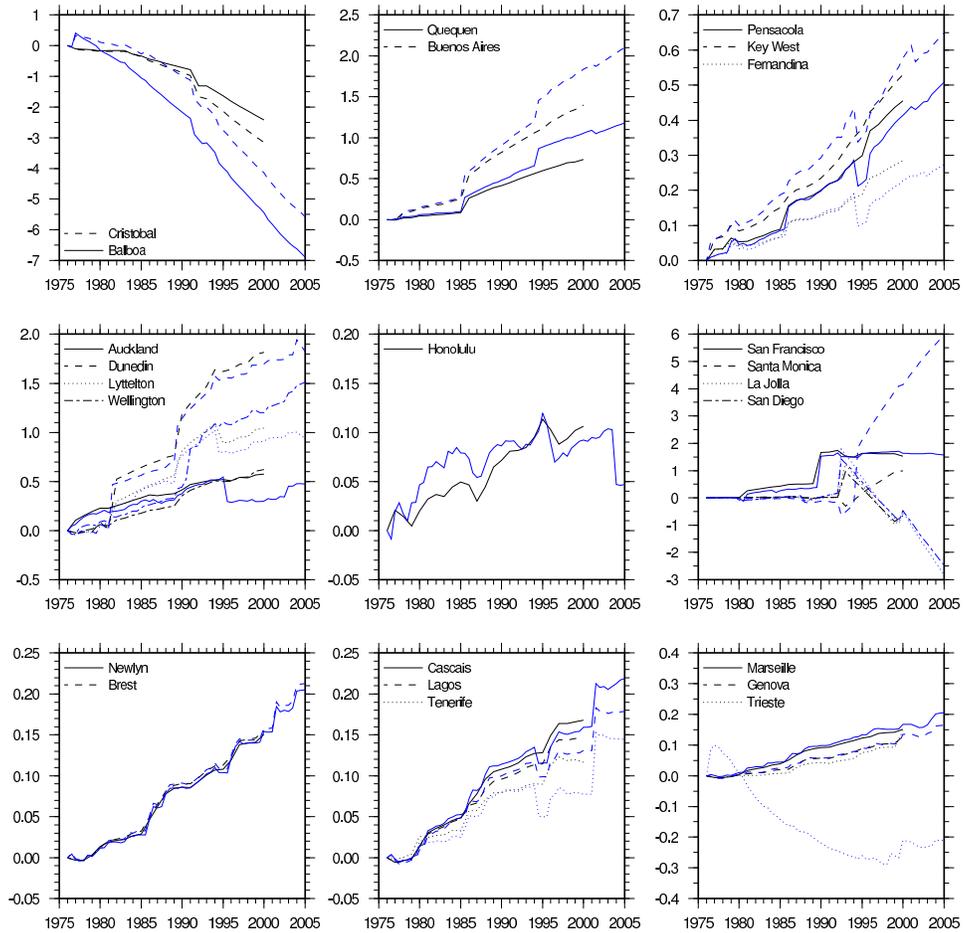}
\end{center}
\caption{Comparison between the expected seismic RSL signal
at the tide-gauge stations considered by \cite{douglas2} computed using
the CMT and PS catalogues in the time-window where the two datasets overlap
(1976--2000). Black lines represent the effect of PS seismicity, blue lines
the effect of CMT seismicity.}
\label{confronto_ps_cmt}
\end{figure}

In figure \ref{confronto_ps_cmt} we compared the RSL time--histories obtained
using PS and CMT catalogues in the period 1976-2000, where the two datasets overlap.
The time--histories obtained with the two seismic catalogues show a correspondence
of the coseismic step-like signals produced by large earthquakes. Since the 
CMT catalogue includes much more events, because of its lower magnitude
threshold, in some cases it may happen that a $M_w < 7$ event (i.e. not
included in PS) located near a tide-gauge
station gives a strong signal in the CMT time-history which is not found in the
PS catalogue. This is, for instance, the case of Trieste tide-gauge, which is affected by
the $M_w = 6.5$ Friuli earthquake occurred on May 6, 1976, which is not reported
by the PS catalogue but gives a large, mostly postseismic, contribution to
the time-histories computed with the CMT.

\subsection{Synthetic analysis}

All the results presented above show a preferentially positive global trend for 
seismically induced sealevel variations. In our previous investigations \citep{rsl01} we
speculated that this behavior might be a consequence of the well-known tendency
of seismic energy release to reduce the oblateness of the Earth 
\citep{chaogross87,chaoetal96,alfspa98}; nevertheless, this
aspect awaits for a better understanding. To this aim, we performed some further analyses.

\begin{figure}
\begin{center}
\noindent\includegraphics[width=30pc]{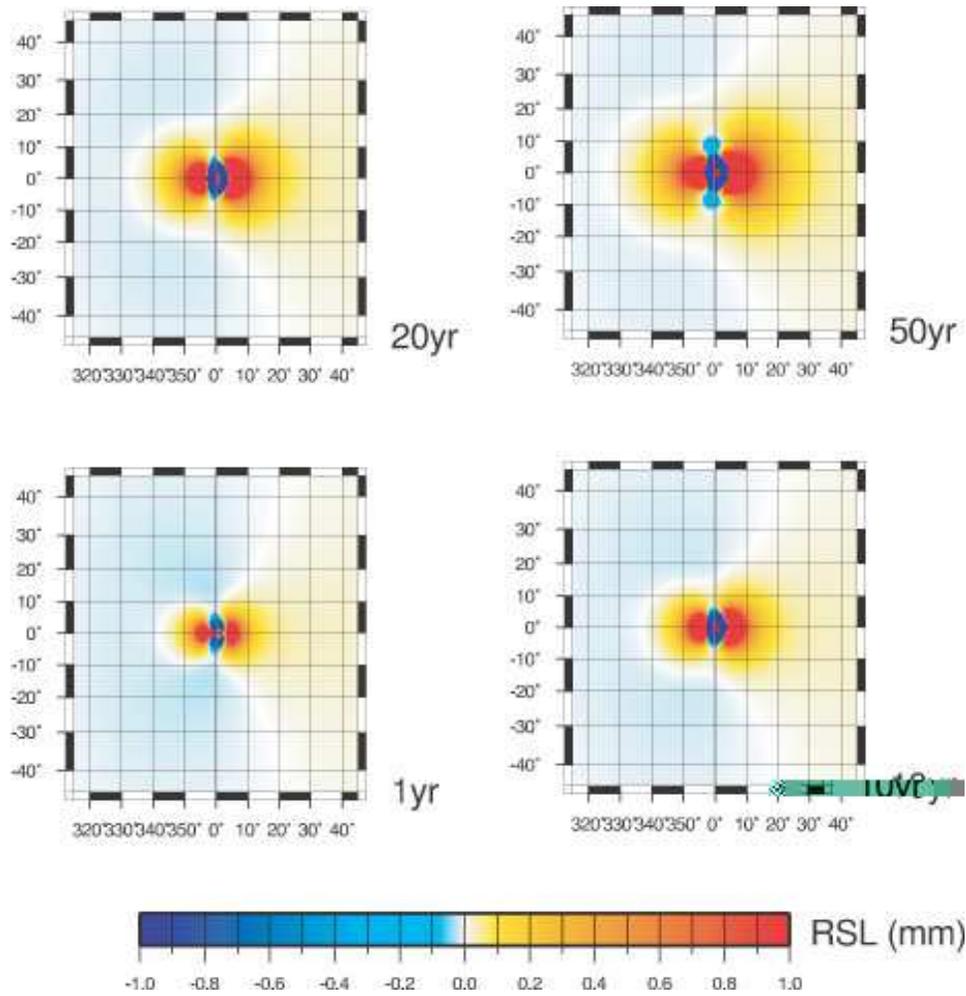}
\end{center}
\caption{Time evolution of the RSL variation induced by a 
$20$ km deep point thrust fault, with strike along the North direction, 
dip angle of $20^\circ$ and seismic moment $M_0=10^{21}$ Nm. 
}
\label{thrustrsl}
\end{figure}

In figure \ref{thrustrsl} we plotted the time evolution of relative sealevel variation induced
by a point thrust fault with $20^\circ$ dip angle, seismic moment $M_0=10^{21}$Nm at a depth
of 20 km. The RSL variation induced by the fault exhibits well separated lobes of positive
and negative variations, whose dimensions are slowly varying with time. Near the epicenter
we find an inner zone of negative variation and an outer zone of positive variation
that expands with time.

\begin{figure}
\begin{center}
\noindent\includegraphics[width=30pc]{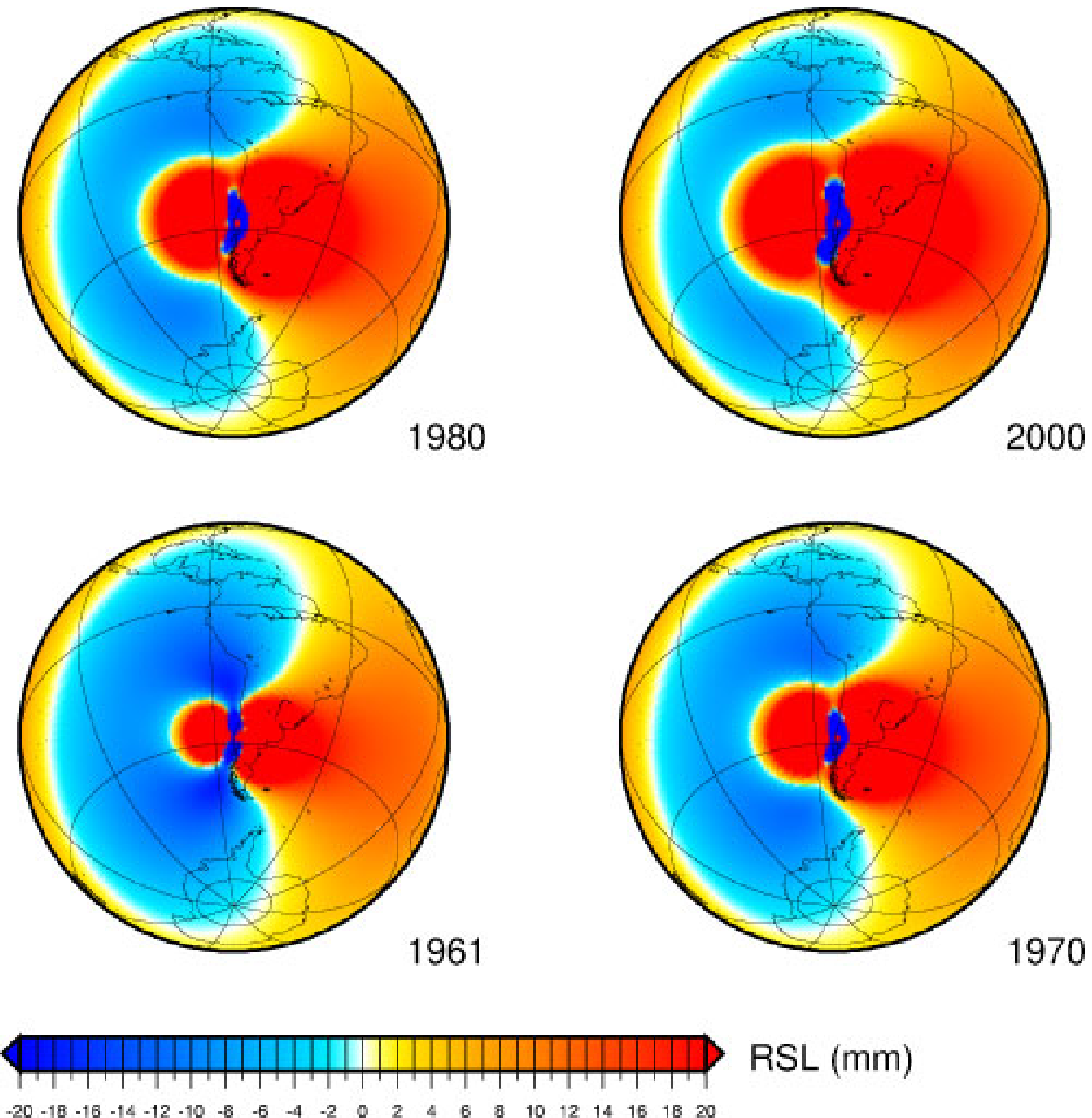}
\end{center}
\caption{Time evolution of $G(t)-u_z(t)$ associated with the 1960 
Chile earthquake. The difference $G-u_z$, when calculated on the
oceanic surface, represents the relative sealevel change.}
\label{cile60}
\end{figure}

Now we focus our attention on the 1960 Chile earthquake (figure \ref{cile60})
that we know to give a major contribution to the total RSL signal 
(see also \cite{rsl01}). If we plot the difference $G(t)-u_z(t)$ on both the oceanic and 
solid surfaces, we can see that the resulting field exhibits
a balanced distribution of areas with positive and negative variations, similar
to the one shown in figure \ref{thrustrsl}. The orientation of the Chile fault
and the geometric properties of the coastlines are such that
the coasts, where the majority of PSMSL tide-gauge stations are located, are in the positive
RSL lobes.

\begin{figure}
\begin{center}
\noindent\includegraphics[width=30pc]{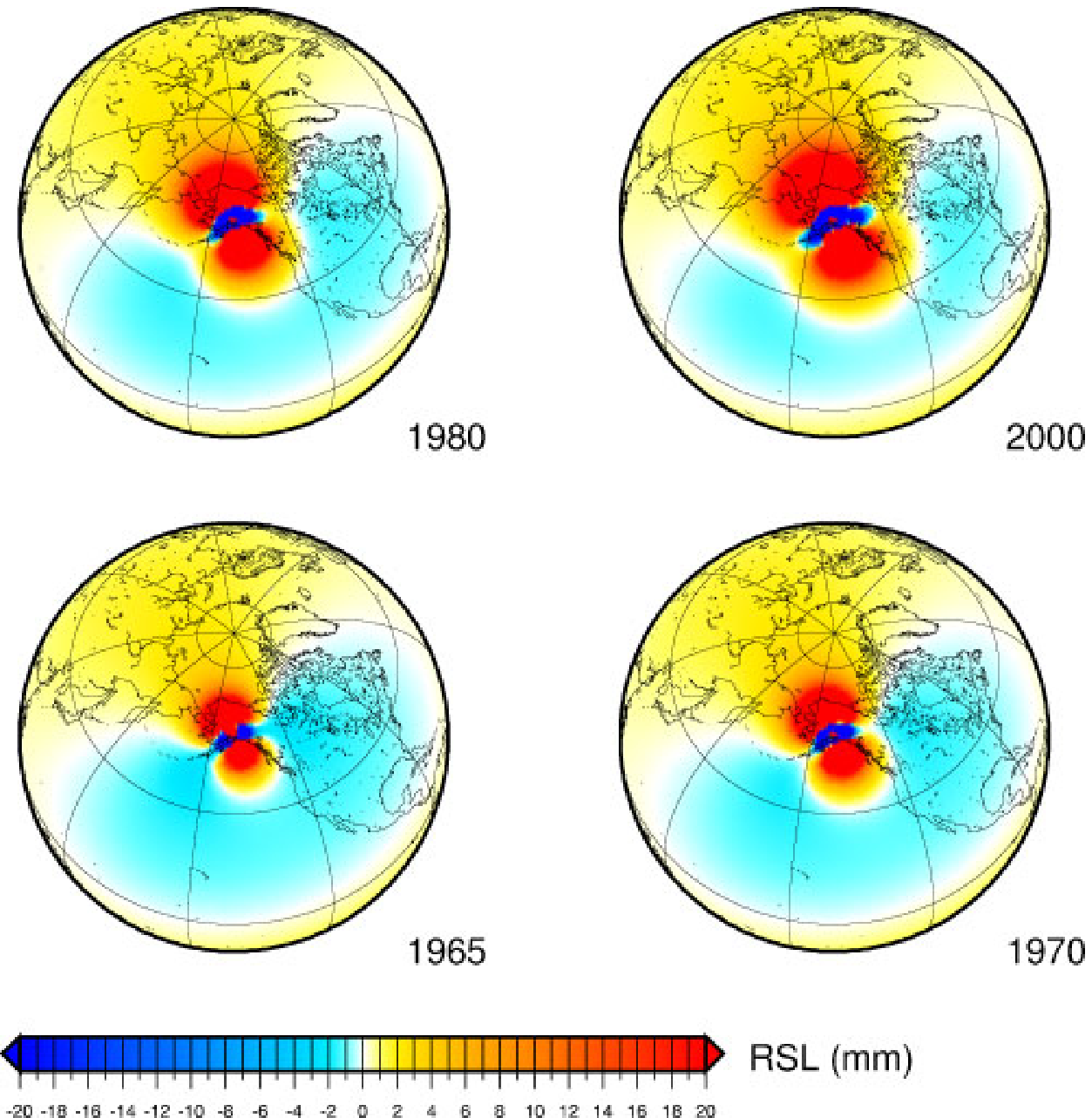}
\end{center}
\caption{Time evolution of $G(t)-u_z(t)$ associated with the 1964 
Alaska earthquake. The difference $G-u_z$, when calculated on the
oceanic surface, represents the relative sealevel change.}
\label{alaska64}
\end{figure}

If we repeat this analysis for the Alaska 1964 (figure \ref{alaska64}) earthquake 
we obtain a similar pattern. Here, the region with negative variations covers
part of the North American eastern coast,
but in the cumulative results (figures \ref{oceani_ps} and \ref{oceani_cmt})
these negative variations are shadowed by the much larger positive signal induced by the 
Chile event in the same area.

Therefore we can see that, in assessing the preferentially positive RSL trend
predicted by our simulations, a major role is played by the reciprocal geometrical properties of the
distribution of tide-gauge stations and orientation of the fault planes of giant earthquakes.

\subsection{Effect of the Sumatra-Andaman earthquake}

On December 26, 2004 an $M_w=9.3$ earthquake stroke the northwestern
coast of the Sumatra island. Current estimates suggest that this is
the second greatest event ever registered. Even if it lies outside the
time window of our analysis, we report some preliminary computation 
about its impact on sealevel.

\begin{figure}
\begin{center}
\noindent\includegraphics[width=20pc,angle=-90]{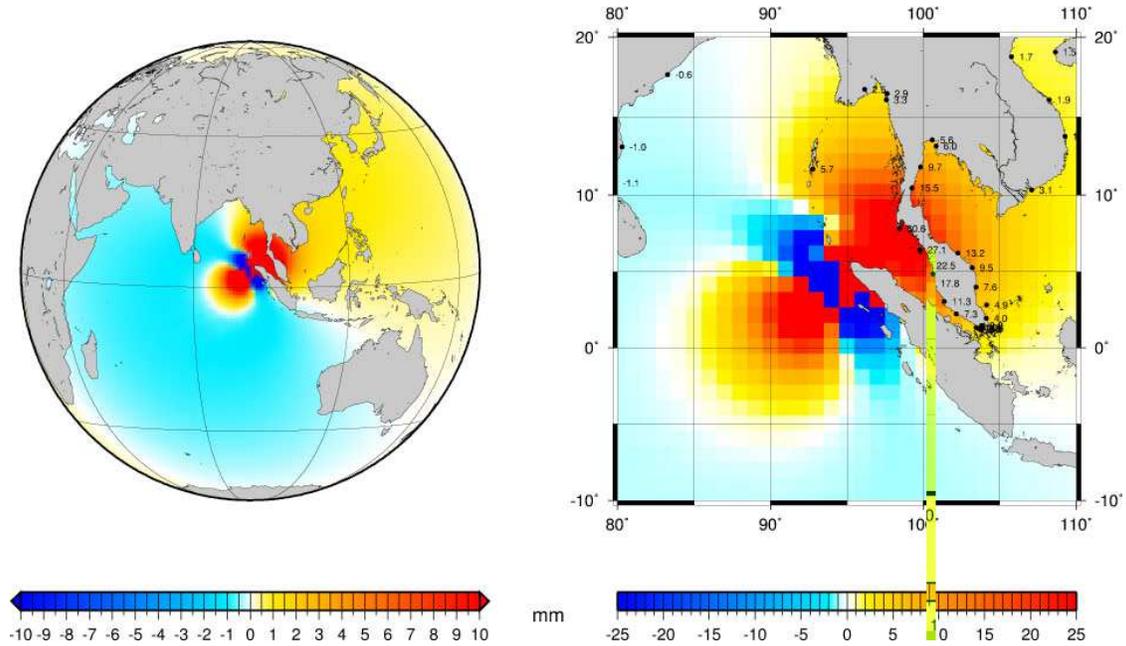}
\end{center}
\caption{Relative sealevel variation associated with the cosesimic
(elastic) deformation and geoid change following the December 26, 2004 
Sumatra-Andaman earthquake. 
In the right panel are reported the locations
of PSMSL tide-gauge stations with the predicted coseismic signal (in mm).
In the computation of both the RSL field and the coseismic offsets on 
PSMSL tide-gauges we assumed the conservation of water volume.}
\label{sumatra}
\end{figure}

In figure \ref{sumatra}, we show the expected relative sealevel variation
following the principal event. The relative sealevel variation remains appreciable
over an extremely large area and should give a strong signal on the
PSMSL tide-gauge stations located in the region.

For this computation, we modeled the seismic source using a preliminary
rupture model (http://www.gps.caltech.edu/~jichen/Earthquake/2004/aceh/aceh.html), 
which assumes a 450km by 180km rupture plane
with 320 degrees strike and 11 degrees dip angles.

Incidentally, we note that for this single event we were able to overcome 
the point dislocation approximation and to adopt a realistic, 
finite seismic source. As a consequence we obtained a precise estimate of 
the geoid correction needed to account for water volume conservation and we ascertained that, as the 
source refinement increases, the geoid correction tends to decrease.

\section{Conclusions}

In this work we further analyzed the effect of seismic activity on sealevel variations.
We improved our computational methods, enabling us to evaluate the effect of 
a larger catalogue of earthquakes (as the CMT) on large grids covering the 
whole oceanic surfaces and to take into account the effects associated with the conservation 
of the total water volume.

Our results confirm the finding of a previous analysis that, on a global scale, 
most of the signal is associated with few giant thrust events. These events can 
induce a sealevel signal on the PSMSL tide gauge stations distributed worldwide 
of about $0.25$ mm/yr. This value is reduced by more than an order of magnitude 
if the effects of the giant thrust earthquakes of the last century are not considered 
(i.e. adopting CMT catalog). Since sealevel rise rates associated with climatological 
factors (water volume increase due to ocean warming and mass increase from ice melting) 
are estimated to be at most $1\div 1.5$ mm/yr \citep{levitus00,milldoug}, the average
contribution to RSL coming from seismic activity is not negligible with respect 
to the climatological factors. Moreover, in regions with strong seismotectonical
activity, the seismic contribution amounts up to several mm/yr representing a major 
contribution to RSL.

In our previous analysis we found that seismicity has an overall tendency to produce
a positive RSL variation, but the reason of this behavior was left unexplained. Now,
we mostly answered to that question: we found that the RSL field induced by
earthquakes has alternating patterns of positive and negative trends, but the geometry
of coastlines, where tidal measurements are taken, is such that the majority of 
tide-gauge stations are located in zones with positive seismic RSL trends. 

A question to be faced in future developments will be the role played by tectonical 
aseismic processes. While our analysis is
far from definitely assessing the role of seismic processes in RSL 
changes, it suggests that the whole tectonic process could be a major non-climatic source of
geoid perturbations and RSL variations; controversial evidence come from 
the results of TOPEX/Poseidon mission that show large deviations from the
mean sealevel trend in regions with important tectonic activity. Altough  
recent investigations have interpreted that satellite-derived sea level trend maps 
purely in term of thermal expansion \citep{willis04}, 
we think that to obtain a complete understanding of sealevel
variations, the seismotectonical contributions can no longer be neglected.

%
%


%
%
%
%
%
%
%

%
%

\begin{acknowledgments}
We thank the Associate Editor (Ctirad Matyska) and two anonymous reviewers
for their incisive comments and helpful suggestions. Partly supported by a 
MIUR-FIRB research grant.
\end{acknowledgments}

%
%

\bibliographystyle{agu04}
\bibliography{biblio,daniele}


%
%
%

%
%
%



%
%
\end{article}
\newpage

%
%
%

\end{document}